\documentclass{aa}
\usepackage{psfig}
\begin{document}

\voffset-1cm
\title{Detection of an X-ray periodicity in the 
Narrow-line Seyfert~1 Galaxy Mrk 766 with {\it XMM-Newton}\thanks{
 Based on observations with XMM-Newton, an ESA Science Mission 
    with instruments and contributions directly funded by ESA Member
    States and the USA (NASA)}
}

\author{Th. Boller \inst{1},
R. Keil \inst{1},
J. Tr\"umper \inst{1},
P.T. O'Brien \inst{2},
J. Reeves  \inst{2}, \and
M. Page \inst{3}
}

\offprints{Th. Boller}

\institute{Max-Planck-Institut f\"ur extraterrestrische Physik,
Postfach 1312, 85741 Garching, Germany\\
\email{bol@mpe.mpg.de} 
\and
X-ray Astronomy Group; Department of Physics and Astronomy; Leicester 
University; Leicester LE1 7RH; U.K.
\and
Mullard Space Science Laboratory, University College London,
Holmbury St. Mary, Dorking, Surrey, RH5 6NT, UK
}

\date{Received October 6, 2000; accepted October 30, 2000 }

\abstract{
We have analyzed the timing properties of the Narrow-line Seyfert 1 galaxy
Mrk 766 observed with {\it XMM-Newton} during the PV phase. 
The source intensity changes by a factor of 1.3 over the 29,000 second
observation. 
If the soft excess is modeled by a black body component, as
indicated by the EPIC pn data, the luminosity of the black body component
scales with its temperature according to $L \sim T^4$. 
This requires a lower limit 'black body size` of
about $\rm 1.3 \cdot 10^{25}\ cm^2$. 
In addition, we report the detection of a strong periodic signal 
with $\rm 2.4 \cdot 10^{-4}\ Hz$.
Simulations of light curves with the observed time sequence and phase
randomized for a red noise spectrum clearly indicate that the periodicity
peak  is intrinsic to the distant AGN.
Furthermore, its existence is confirmed by the EPIC MOS and RGS data.
The spectral fitting results show that the black body temperature and the
absorption by neutral hydrogen remain 
constant during the periodic oscillations.
This observational fact tends to rule out models in which the intensity
changes are due to hot spots orbiting the central black hole.
Precession according to the Bardeen-Petterson effect 
or instabilities in the inner accretion disk
may 
provide explanations for the periodic signal. 
}

\maketitle

\markboth{Th. Boller et al.:Detection of an X-ray periodicity in Mrk 766}{Th. Boller et al.:Detection of an X-ray periodicity in Mrk 766}
\section{Introduction}

Mrk 766 is a nearby ($z$ =  0.013) and X-ray  bright ($\rm \sim 10^{-11}\ erg\ s^{-1}$) Narrow-Line Seyfert~1 galaxy 
(Walter \& Fink 1993, Boller et al. 1996, Leighly et al. 1996, Leighly 1999).
Due to the extreme spectral and X-ray properties found
in Narrow-Line Seyfert 1 galaxies and due to its extreme X-ray brightness,
Mrk 766 was proposed for the PV phase for observations with the X-ray
satellite {\it XMM-Newton}. The high throughput of {\it XMM-Newton}
compared to previous X-ray missions allows precise studies of the
spectral properties, e.g. the shape of the soft excess and of the
transition region between the soft excess and the power-law component.
In addition, RGS observations allow to perform X-ray spectroscopy 
of accretion disk lines (e.g. Ross et al. 1999) as well as 
emission and absorption features originating from the broad- and narrow-line
region. 
The {\it XMM-Newton} spectral properties of Mrk 766 are discussed in detail by 
Page et al. (2001) and Pounds et al. (2001). 
The long-period orbit of XMM allows to measure the variability
power spectra of ultrasoft Narrow-Line Seyfert 1 galaxies far better 
than has yet been possible before and allow to constrain nonlinear variability, to search for
spectral variability and to
search for any quasi-periodic oscillations (e.g. Papadakis \& Lawrence 
1995 and references therein).
In this paper we  concentrate on the timing properties of Mrk 766 and
present the discovery of an X-ray periodicity in Mrk 766. 
The object resembles IRAS 18325$-$5926 (Iwasawa et al. 1998), where
a 58000 second periodicity in the 0.5$-$10 keV {\it ASCA} energy band was found.

\begin{figure}
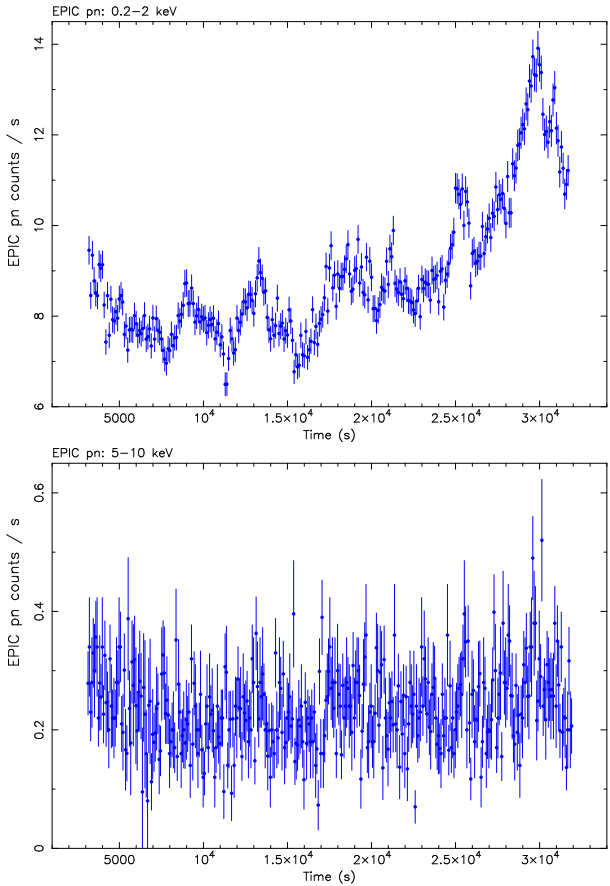

\centering
\centerline{\psfig{figure=xmm30_f1.eps,width=8.0cm,angle=-90,clip=}}
\centerline{\psfig{figure=xmm30_f2.eps,width=8.0cm,angle=-90,clip=}}
      \caption{
XMM Newton EPIC pn light curve of Mrk 766 in the 0.2--2 and 5--10 keV 
energy band, respectively.
}
\end{figure}

\begin{figure}
\centering
\psfig{figure=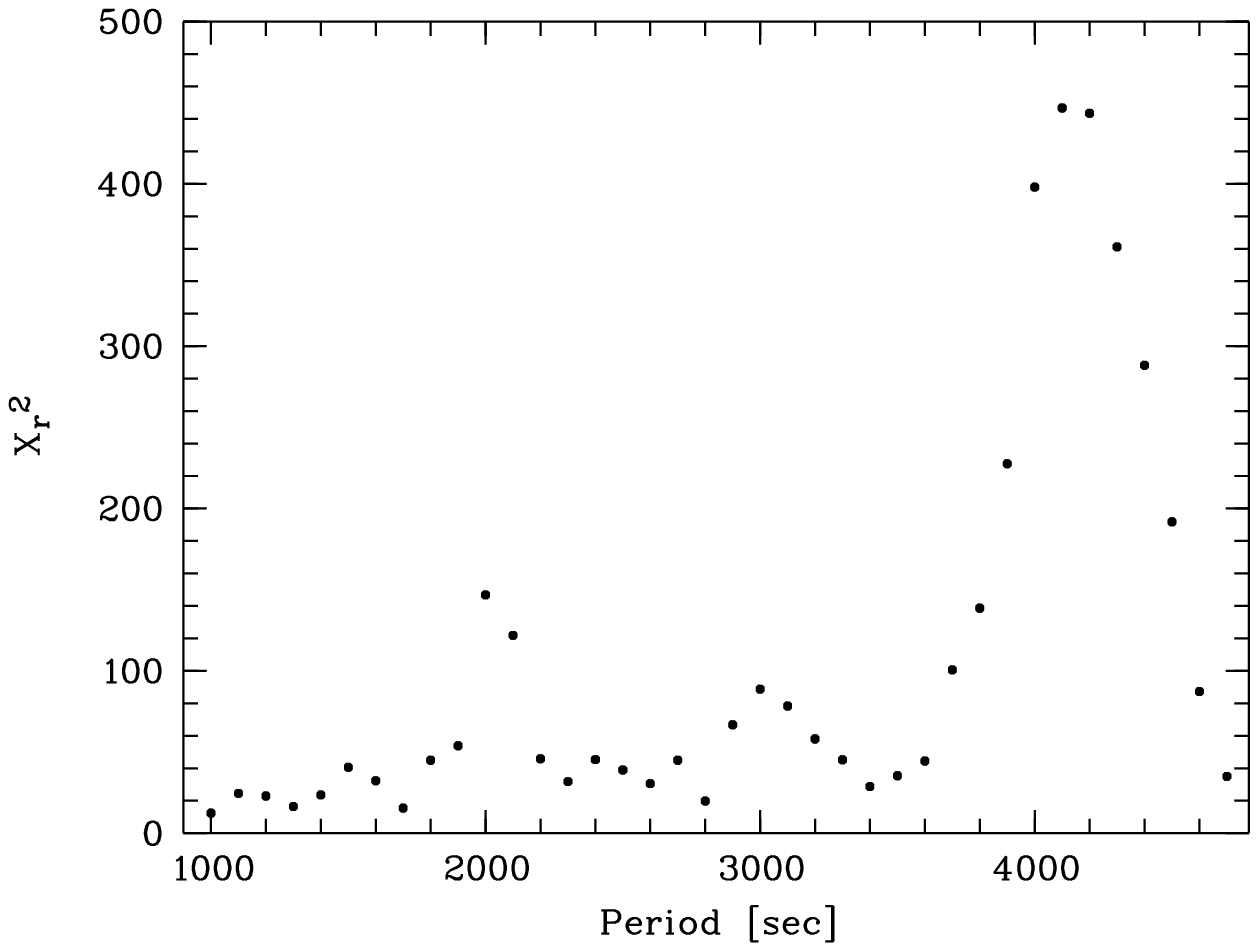,width=8.0cm,height=5cm,clip=}
\vskip 0.2cm
\psfig{figure=xmm30_f4.eps,width=7.94cm,height=5cm,angle=-90,clip=}
\vskip 0.2cm
\psfig{figure=xmm30_f5.eps,width=8.45cm,height=5cm,angle=-90,clip=}
      \caption{
Upper panel: Reduced $\rm \chi^2$ versus the folding period for 
the 0.2-10 keV energy
range. 
Middle panel:
Folded 0.2--2 keV light curve with the best fitting period of 4200 seconds. 
Lower panel: Power spectrum for Mrk 766 in the 0.2-10 keV energy band.
}
\end{figure}

\section{The {\it XMM-Newton} PV observations on Mrk 766 and data reduction}

The Narrow-line Seyfert 1 galaxy Mrk 766 was observed during the
performance and calibration phase in revolution 0082 at May 20, 2000.
The EPIC pn camera was operated  in the small window mode implying a frame
read out time of 30 msec so that pile-up is not significant. The full frame
mode was used for the EPIC MOS cameras and the RGS cameras were operated
in the standard spectroscopy+Q mode. In total 13 exposures were taken
with the OM using the UVW1, UVW2 and UVM2 filter. 
At the beginning of the observations the EPIC cameras were blocked for
20 ksec due to high particle background. At the end of the observations
all instruments suffered from  data gaps. The exposure time used 
for the spectral and timing analysis presented in this paper is 29000 seconds.
For the data reduction and data analysis the standard SAS software
packages were used. The spectral fitting and light curve analysis was
performed using {\tt XSPEC version 10.00} and {\tt XRONOS version 4.0},
respectively. 

\subsection{Detection of a periodicity in the X-ray light curve of Mrk 766}

\subsubsection{EPIC pn data analysis} 

The EPIC pn 0.2--2 keV and 5--10 keV light curves of Mrk 766 are shown in Figure 1.
The most striking feature is the presence of a periodic signal in 
the 0.2--2 keV energy band
superimposed on a longer-term count rate increase.
Starting at 5000 seconds, the most significant intensity
peaks are separated by 4050, 4210, 4210, 3500, 4050 and 4380 seconds.
In addition, lower amplitude variations separated by about 2100 seconds
are visible (cf. the  peak emission at about 27500 and 31000 seconds).
The 5--10 keV light curve shows no significant count rate oscillations.
We have folded the light curve with different periods, ranging
from 1000 and 5000 seconds, and have determined the corresponding 
$\chi_{\rm r}^{2}$ value. 
The best fitting period is found at 4200 seconds (cf.
upper panel of Figure 2).
The corresponding folded light curve is given in the middle panel of Figure 2. 
The power spectrum (lower panel) confirms the presence
of a strong periodic signal with $\rm 2.4 \cdot 10^{-4}\ Hz$. 
The second highest value is found at twice that frequency.

\subsubsection{Comparison with EPIC MOS and RGS data}
The MOS and RGS data are basically
consistent with the EPIC pn results showing a peak near 4200 seconds.
In Figure 3 we show the folded light curves for the MOS-1 and RGS-1 detectors.
Although the statistical errors are larger in the MOS and RGS data, the
period of 4200~seconds is also confirmed by the MOS and RGS data.
\begin{figure}
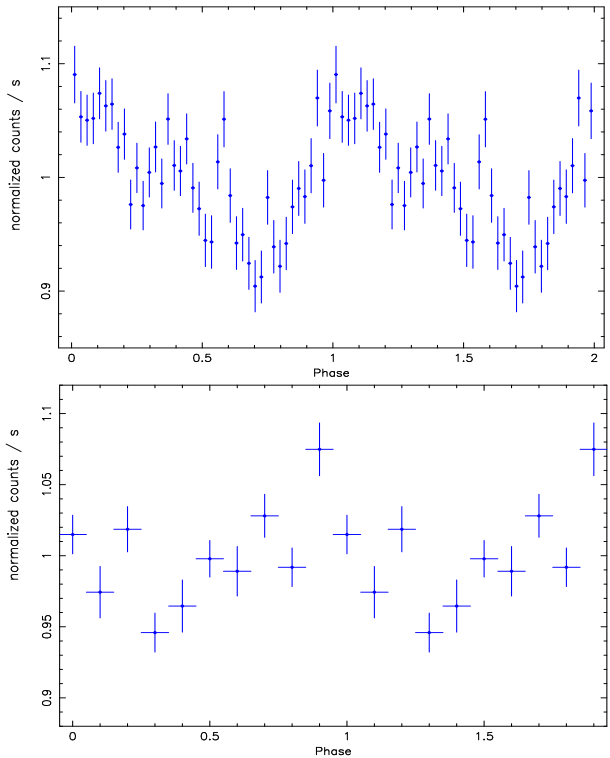

\centerline{\psfig{figure=xmm30_f6.eps,width=8.0cm,height=5cm,angle=-90,clip=}}
\centerline{\psfig{figure=xmm30_f7.eps,width=8.0cm,height=5cm,angle=-90,clip=}}
      \caption{
Folded MOS-1 light (upper panel) and RGS-1 light (lower panel) 
curves using the best fitting
period of 4200 seconds. The folded RGS light curve covers a longer time scale
compared to the EPIC data, therefore both light curves are not in phase. 
}
\end{figure}

\subsubsection{Monte-Carlo Simulations on red noise light curves}
\begin{figure}[htp]
\centerline{\psfig{figure=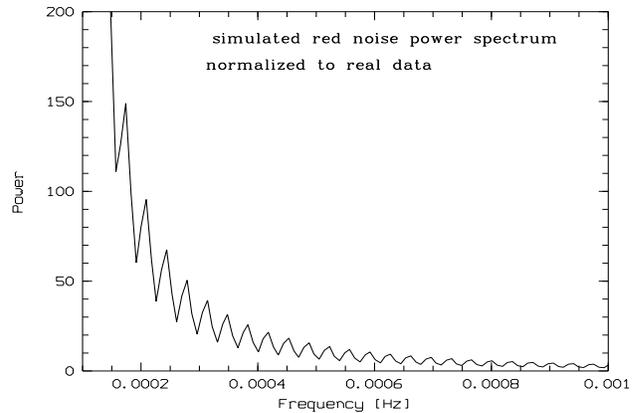,width=8.7cm,height=5.5cm,clip=}}
      \caption{Simulated red noise power spectrum of Mrk 766, normalized
to the real data set. 
}
\end{figure}
We made simulations of light curves with the observed time sequence
and phase randomized for a red noise $f^{-1}$ power spectrum (cf. Figure 4). 
The periodic signals detected in the X-ray
light curve of Mrk 766 at 4200 and 2100 seconds  
(cf. lower panel of Figure 2) far exceed the power seen in the 
simulated power spectrum. 
Therefore we conclude that the periodicity peaks are not caused by red noise and
are intrinsic to the Narrow-Line Seyfert 1 galaxy
Mrk 766.

\subsection{EPIC pn spectral analysis}
\begin{figure}[htp]
\centerline{\psfig{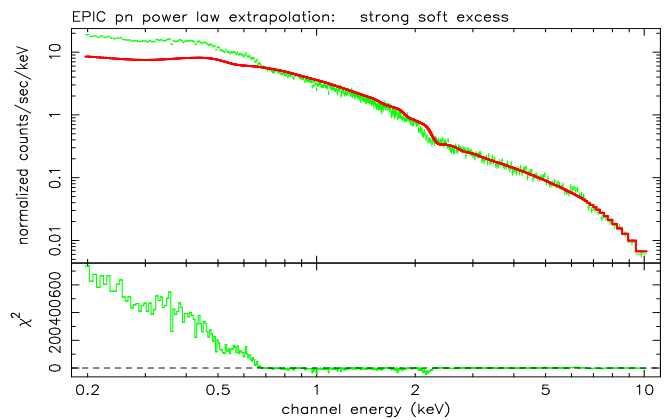}}
      \caption{Extrapolation of the power-law component down 
to soft X-ray energies. A strong soft excess component is detected
below about 0.7 keV.
}
\end{figure}
\begin{figure}
\centerline{\psfig{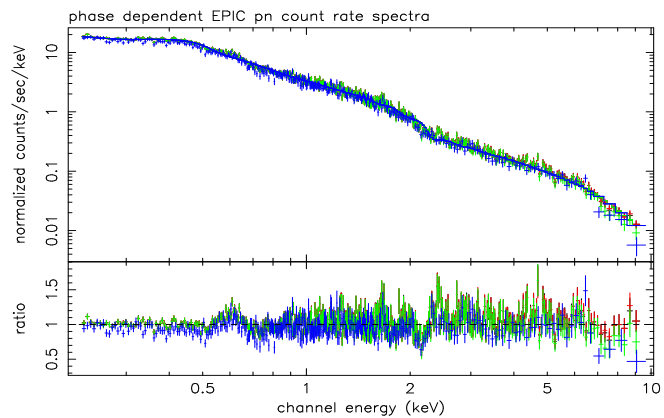}}
      \caption{Black body plus power-law fit to EPIC pn spectra selected
from four different phase intervals. No significant changes in the
black body temperature and the absorption by neutral hydrogen is detected.
}
\end{figure}
The 0.2--10 keV EPIC pn spectrum of Mrk 766 is well described by
a power law model with a photon index of $\Gamma = 2.11 \pm 0.10$
dominating the 1 to 10 keV flux, in combination
with a strong soft excess component which dominates the X-ray flux below 
about 1 keV.
In Figure 5 we show the EPIC pn spectrum where the power-law photon index
is extrapolated down to 0.2 keV.  
\begin{figure*}[htp]
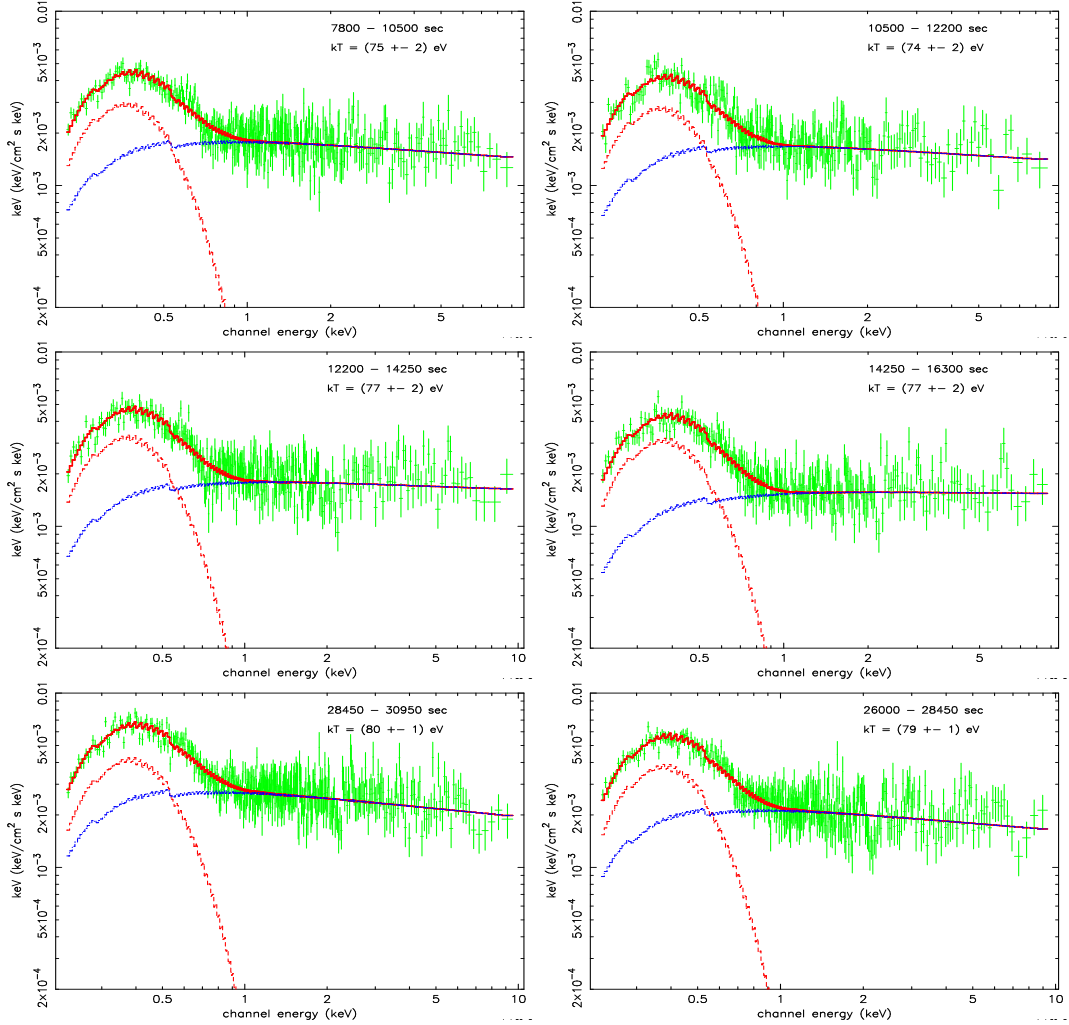

\mbox{
\centerline{\psfig{figure=xmm30_f11.eps,width=7cm,height=4.5cm,angle=-90,clip=}
\psfig{figure=xmm30_f12.eps,width=7cm,height=4.5cm,angle=-90,clip=}}}
\mbox{
\centerline{\psfig{figure=xmm30_f13.eps,width=7cm,height=4.5cm,angle=-90,clip=}
\psfig{figure=xmm30_f14.eps,width=7cm,height=4.5cm,angle=-90,clip=}}}
\mbox{
\centerline{\psfig{figure=xmm30_f15.eps,width=7cm,height=4.5cm,angle=-90,clip=}
\psfig{figure=xmm30_f16.eps,width=7cm,height=4.5cm,angle=-90,clip=}}
}
      \caption{Black body and power-law fit to the EPIC pn spectra of
Mrk 766 for different time intervals.
Upper two panels: Spectra selected for the high (left panel)
and the low (right panel) flux state for the 
periodicity peak centered at 9000 seconds (corresponding to the time
given in Figure 1.)
The temperature of the black body component remains constant within the errors. 
Middle and lower panels: Same as above for the periodicity peaks centered at 13220 and 29700 
seconds. 
}
\end{figure*}

The EPIC pn spectrum clearly indicates the
presence of a strong soft excess component superimposed on an underlying
power-law component. Between 0.2--0.7 keV the soft excess component 
contains about 40 per cent more flux than the power law component.
The soft excess component is well described by a black body spectrum.
Signatures for a strong
warm absorber component are not detected in the EPIC pn spectrum. 
Although spectral
residua between about 0.7--0.8 keV seem to be present in the 
RGS spectrum (Page et al. 2001),
these residua cannot explain the spectral kink detected at about 1 keV
in the EPIC pn spectrum (cf. Figure 7). 
We conclude that 
absorption due to a warm absorber, if present, is small compared to the soft
excess emission and is unable to produce the spectral kink at around 1 keV.
Therefore a combination of a black body component and an underlying
power law is used for the spectral fitting described in the following. 
In Figure 6 we show EPIC pn count rate spectra selected according to
4 different phase intervals  (middle panel of Figure 2). The count rate
spectra were obtained from the phase intervals ranging from 
0$-$0.2 (phase A), 
0.2$-$0.55 (phase B), 
0.55$-$0.85 (phase C), and
0.85$-$1.0 (phase D), respectively. 
These phase intervals correspond to different
count rate states during the X-ray flux oscillations. 
The spectral fitting results indicate that the black body temperature
remains constant within the errors for the 4 phase intervals.
The black body temperature is 
76 $\rm \pm$ 1 eV (phase A), 
76 $\rm \pm$ 1 eV (phase B),
75 $\rm \pm$ 1 eV (phase C) and
76 $\rm \pm$ 1 eV (phase D).
The absorption by neutral hydrogen is consistent with 
the Galactic value of $N_{\rm H, Gal} \rm  = 1.8 \cdot 10^{20}\  {cm^{-2}}$
(Dickey and Lockman 1990) for the four phase intervals.
In addition, we show in Figure 7 the spectral fitting
results for three different time intervals of the X-ray
light curve shown in Figure 1. The spectra were selected for the
high and low flux states for 3 periodicity peaks centered at 9000,
13200 and 29700 seconds, respectively. We find that the black body 
temperature remains constant within the errors during the X-ray
oscillations for each periodicity peak.

\begin{figure}
\centerline{\psfig{figure=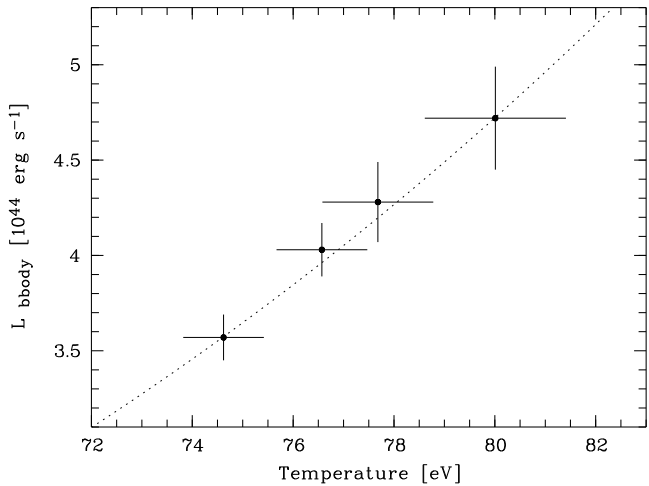,width=8.7cm,clip=}}
      \caption{Black body temperature versus the bolometric black body 
luminosity for 4 different time intervals selected from the light curve shown
in Figure 1. The temperature increases with the source luminosity according
to a black body emission law.
}
\end{figure}
To study the long-term variability we have divided the light curve
shown in Figure 1 into 4 different time intervals, ranging from
7000$-$17100, 17100$-$24500, 24500$-$28500, and 
28500$-$31000 seconds, respectively. For each time interval we have
performed the combined power law plus black body fit and have
determined the temperature and the
bolometric luminosity of the black body component. 
With Figure 8 we show that the black body temperature increases from
74.6 to 80.1 eV and that bolometric luminosity increases from about 
$\rm 3.57 \cdot 10^{44}\ \rm erg\ s^{-1}$
to $\rm 4.72 \cdot 10^{44}\ \rm erg\ s^{-1}$ (for $H_{\rm 0} = 
\rm 70\ km\ s^{-1}\ Mpc^{-1}$ and $q_{\rm 0} = \rm \frac{1}{2}$).

\section{Discussion}

We
first discuss the 'long-term` increase observed in the light curve
(cf. Figure 1) which is characterized by a correlation between flux and 
temperature of the soft component. To our knowledge such a behavior has not
been found in AGN spectra before. The correlation between the bolometric
luminosity and the spectral temperature is close to the 
$L_{\rm bb} \sim T^4$ law, consistent with a radiating surface
remaining constant during the variations.
The projected size of the black body radiating
disk surface can be estimated 
using the luminosities
and temperatures given in Figure 8. We find that this projected disk size
is
$\rm 1.3 \cdot 10^{25}\ cm^2$
which is 
rather a lower limit to the physical size
of the emission region, because 
we have neglected a possible inclination of the disk,
gravitational redshift effects and deviations
from a black body due to scattering effects.
We note that this area estimate requires a black hole mass of at least
a few $\rm 10^5$ solar masses.

The periodic signal may be a quasi-periodic oscillation (QPO).
Investigations of QPOs in  galactic binary systems show a wide range
of frequencies between about 0.1 and 1200 Hz (Psaltis et al. 1999).  
However, we note that for the galactic black hole microquasars GRS 1915+105
and GRO 1655-40 narrow frequencies of roughly 50 -- 300 Hz have been
detected (Morgan et al. 1997, Remillard et al. 1999). 
The ratio between these QPO frequencies and the strongest peak frequency  
found in the data of Mrk 766 lies between $\rm 10^{5-6}$
roughly consistent with 
the expected mass ratio of these objects.
Specifically, QPOs could be produced by instabilities in the inner part of the
accretion disk or by pulsating accretion if the rate is close to the Eddington
limit. In these cases the period is expected to be of the order of the radial
drift time scale (`instability time scale')
$$
\tau  \sim \epsilon \cdot \alpha^{-1} \left(\frac{r}{R_{\rm S}}\right)^{\frac{3}{2}} \cdot 
\frac{R_{\rm S}}{c}   \eqno(1)
$$ 
where $\epsilon \sim$ 5 to 10 (Sunyaev 2000).
We note that $\tau$ is
of the order of a few thousand seconds for $\alpha \sim \rm 0.1$,
$r \sim 6\ R_{\rm S}$ and $M \sim \rm  10^6$ solar masses.

If the
periodic signal arises from X-ray hot spots orbiting the black
hole (e.g. Sunyaev 1973; Guilbert, Fabian \& Rees 1983)
we expect a variation in temperature.  Using the
radius dependent orbital velocity as given by Abramowicz et
al. (1996), the boost model of Rybicki \& Lightman
(1979) and assuming an inclination angle of 45 degrees we find a
temperature variation by a factor of about 1.6 and 1.4 at 3 and 4 Schwarzschild
radii, respectively.
As shown in Figure 7 the black body temperature remains constant
within a few percent during the periodic variations 
requiring an
inclination angle of the accretion disk of more than 85 degrees.
For 85 degrees the temperature change according to the moving black body
field is about a factor of 1.1 at 3 Schwarzschild radii, 
still inconsistent with the observations.
Even for a nearly face-on geometry with an inclination angle of 89 degrees
temperature changes by a factor of 1.06 are expected from the hot spot model.

Another possibility is that the observed variability is due to disk
precession according to Bardeen and Petterson (1975).
This causes the inner disk to settle into the plane of the black
hole equator up to the so called transition radius, whereas the outer
parts remain in their original orientation. The disk precession takes place in a ring
which is in between.
Nelson \& Papaloizou (2000) found that the transition radius
is of the order of about 10 to 30 gravitational radii.
Since the precession of the disk is a geometrical effect it does not
lead to spectral changes which is consistent with the observations.
The absence of significant absorption by neutral gas above the galactic
value shows that shadowing by the outer parts of the accretion disk is not
responsible for the observed periodicity. The amplitude of the variations
of about 7 per cent (see Figure 2) requires
a precession angle dependent on the disk inclination 
according to $\Delta i = \frac{{\rm 0.07}}{{\rm cot}\ i}$.
For instance, for a disk inclination of 90 (45) degrees the precession angle
is 32 (4) degrees.
The angular shape of the light curve may reflect the expected disk warp. 
The second peak at 2100 seconds in the power spectrum and the possible third
one at about 3000 seconds may be related to the warp as well. 
The precessing frequency produced by the Bardeen-Petterson effect
depends on the disk radius $r$ as
$$
\omega =  \frac{2G}{c^2} J r^{-3} \eqno(2)
$$
where $J$ is the angular momentum of the black hole.
The maximum angular momentum of a Kerr black hole is
$$
 J_{\rm max} = G\ M^2\ c^{-1} \eqno(3)
$$ 
For the ratio between $J$ and $J_{\rm max}$ we get

$$
\frac{J}{J_{\rm max}} = \frac{\omega\ c^3}{2 G^2} \cdot \frac{r^3}{M^2} \eqno(4)
$$
Equ. 4 can be used to estimate the lower limit of the black hole mass
in Mrk 766:
Taking $\omega \rm = 1.5 \cdot 10^{-3}$ and  the 
black body disk radius of $\rm 2 \cdot 10^{12}\ cm$ as an rough estimate
for $r$, we find that 
a black hole mass of $M \rm > 3 \cdot 10^6$ solar masses is required not to violate
the angular momentum limit.

Presently, we are unable to differentiate  between
a strictly periodic behavior and quasi-periodic oscillations with high 
precision.
We also need better data for the light curve and the power spectrum
in order to constrain the models. In this context 
the 
forthcoming 150 ksec {\it XMM-Newton} observation will be most useful.

\section{Summary}

During the long-term
variations of Mrk 766 the fitted black body temperature and luminosity
of the soft component correlates as expected for a black body 
($L \sim T^4$) requiring a lower limit of the black body disk size
of about $\rm 10^{25}\ cm^2$.
In addition, we have detected a strong periodic signal in the black body 
component with 
4200 seconds.
This periodicity may be of the QPO type resulting from instabilities
in the inner accretion disk around the black hole.
X-ray periodicity due to hot spots  orbiting the black hole does not
provide a plausible explanation, except for the case the disk 
is observed nearly face on ($i$~$\ge$~85 degrees). 
As no significant periodic changes of the black body temperature are detected 
precession may
provide another explanation for the X-ray periodicity of
Mrk~766. 
If the Bardeen-Petterson effect is responsible for the observed periodicity,
the required size of the emission region calls for a 
black hole mass of $M \rm > 3 \cdot 10^6$ solar masses.

\acknowledgements
It is a pleasure to acknowledge the efforts of the SOC and SSC teams
in making the observations possible and for developing the SAS software
package used to reduce the data. 
The XMM - Newton project is supported by the Bundesministerium f\"ur
    Bildung und Forschung / Deutsches Zentrum f\"ur Luft- und Raumfahrt 
    (BMBF/DLR), the Max-Planck Society and the Heidenhain-Stiftung.
We are grateful to the referee, K. Iwasawa, for very useful comments 
which helped
us to improve the paper substantially.
We are indebted to B. Aschenbach for fruitful discussions on the
scientific content of this paper.
TB thanks A.C. Fabian for using his software code on the red noise simulations.

\end{document}